\documentclass[english,twocolumn]{revtex4}
\usepackage[T1]{fontenc}
\usepackage[latin1]{inputenc}
\usepackage{float}
\usepackage{graphicx}
\usepackage{amsmath}
\usepackage{amssymb}

\makeatletter
\usepackage{babel}
\makeatother
\begin{document}

\title{Visibility of graphene flakes on a dielectric substrate}

\author{D. S. L. Abergel, A. Russell and Vladimir I. Fal'ko}

\affiliation{Physics Department, Lancaster University, Lancaster, LA1 4YB, UK}

\date{\today}

\begin{abstract}
We model the optical visibility of monolayer and bilayer graphene
deposited on a SiO$_{2}$/Si substrate or thermally annealed on the
surface of SiC. Visibility is much stonger in reflection than in transmission,
reaching the optimum conditions when the bare substrate transmits
light resonantly. In the optical range of frequencies a bilayer is
approximately twice as visible as a monolayer thereby making the two
types of graphene distinguishable from each other. 
\end{abstract}
\maketitle
Monolayer graphene is a single two-dimensional honeycomb lattice of
carbon atoms. Although the first graphene-based structures were only
recently fabricated \cite{bib:Novoselov-science} they have quickly
become the subject of an extensive research effort \cite{bib:novoselov-nature,bib:Zhang-prl,bib:Geim-natmat}.
Monolayer graphene is a zero-gap semiconductor with a Dirac-like dispersion
of chiral quasiparticles near the $K$ points of the hexagonal first
Brillouin zone \cite{bib:Dress-Ando}. Bilayer graphene is a pair
of graphene sheets with the Bernal (AB) stacking arrangement. In the
low-energy spectrum of this material \cite{bib:McCann-PRL} the conduction
and valence bands both consist of two quadratic branches split by
the inter-layer coupling $\gamma_{1}$. Measurements of the quantum
Hall effect \cite{bib:novoselov-nature,bib:Novoselov-science,bib:Novoselov-natphys}
and ARPES experiments \cite{bib:SiC-Ohta} have confirmed that these
are the low-energy band structures of these materials.

\begin{figure}[tb]
	\centering
	\includegraphics{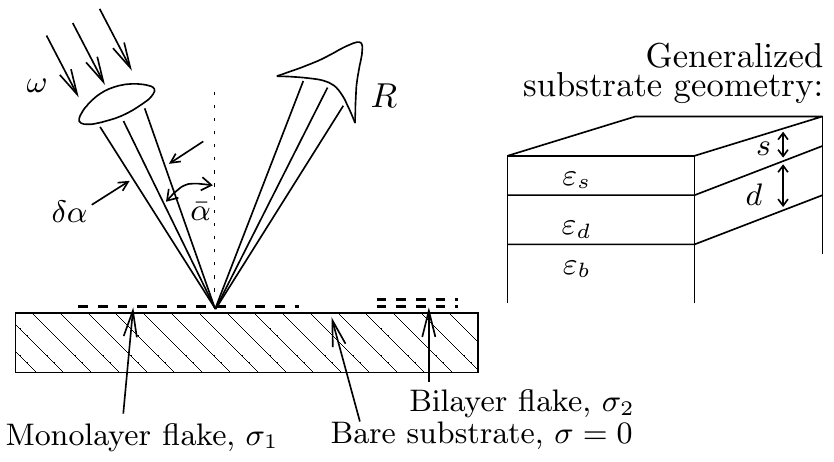}
	\caption{Geometrical configuration for detection of graphene on a
	substrate.  A light beam is focused on a small spot which is scanned
	along the surface. The calculations presented below show how to
	optimise conditions for visibility of atomically thin graphitic
	flakes. \label{fig:exp_setup}}
\end{figure}

The widespread microcleavage technique used to fabricate graphene-based
devices requires a visual inspection of the substrate \cite{bib:Novoselov-science}
to find flakes of one or two layers thickness. In this Letter, we
aim to determine the optimum conditions for making these flakes optically
visible when they are deposited on various substrates. The parameters
at one's disposal (see Fig. \ref{fig:exp_setup}) are the frequency
$\omega$, angle $\bar{\alpha}$ and aperture $\delta\alpha$ of the
focused incident radiation, as well as the thicknesses of the various
layers of the underlying dielectric materials.

Below we calculate the reflection of non-polarized incident light
taking the geometry of the substrate into account with suitable boundary
conditions at each of the interfaces between materials, appropriate
frequency-dependent dielectric functions $\varepsilon(\omega)$ for
each layer, and $\mu=1$. Throughout the calculation, we use the data
\cite{bib:permittivity-details} available in the existing literature
for the dispersion of the permittivity of silicon \cite{bib:si-permittivity},
silicon oxide \cite{bib:SiO2-permittivity} and silicon carbide \cite{bib:SiC-permittivity}.
With reference to Fig. \ref{fig:exp_setup}, we analyze the reflection
$R$ of light from a substrate with a flake on it and compare this
to the reflection $R_{0}$ of a bare (graphene-free) substrate. The
optical visibility of a flake is then determined as the contrast between
two such parts of the sample studied using a monochromatic light source:
\begin{equation}
V_{R}=(R-R_{0})/R_{0}.\label{eq:visdef}\end{equation}

The scattering of light is analyzed using the electromagnetic wave
equations in vacuum and dielectric media and the standard boundary
conditions at interfaces between different materials, \begin{gather}
\vec{E}_{1}^{\parallel}=\vec{E}_{2}^{\parallel},\quad\vec{D}_{1}^{\perp}=\vec{D}_{2}^{\perp},\quad\vec{B}_{1}^{\parallel}-\vec{B}_{2}^{\parallel}=\sigma(\omega)\vec{E}\times\vec{n}.\end{gather}
 The superscripts $\parallel$ and $\perp$ stand for the components
of the field parallel and perpendicular to the interface respectively,
$\vec{n}$ is the unit vector normal to the interface, the subscript
1 (2) denotes the field below (above) the interface, and $\sigma(\omega)$
is the frequency-dependent conductivity of a graphene flake and $\vec{D}=\epsilon(\omega)\vec{E}$.
One more boundary condition (on the perpendicular components of $\vec{H}$)
duplicates Snell's law.

Having in mind an optical setup used to locate a small flake, we consider
a beam of light focused by a lens, so that the light in the beam arrives
at the substrate surface with some aperture $\delta\alpha$ (see Fig.
\ref{fig:exp_setup}). Therefore the measurable reflectance to be
used in Eq. (\ref{eq:visdef}) is \begin{equation}
R(\bar{\alpha},\delta\alpha)=\int d\Omega_{\vec{k}}R(\vec{k})P(\vec{k}),\label{eq:R-integral}\end{equation}
where $P(\vec{k})$ characterises the spread of the beam over the
solid angle of the aperture $\delta\alpha$ around $\bar{\alpha}$,
$\vec{k}=\frac{\omega}{c}(\sin\alpha,0,-\cos\alpha)$ is the wave
vector of the incident ray of light, and $R(\vec{k})$ is the reflection
coefficient for a plane wave with this wave vector. Below we assume
that the beam is equally dense at all angles within an aperture of
$\delta\alpha$ around $\bar{\alpha}$.

To describe the conductivity of graphene, we follow the method used
in Refs \cite{bib:Falkovsky-cond,bib:Nilsson-PRL} taking into account
the split bands formed in the bilayer \cite{bib:McCann-PRL}. At low
temperatures the result for the monolayer which takes into account
the transition between the valence and conduction bands in the Dirac
spectrum is $\sigma_{1}=e^{2}/4\hbar$ (with a negligible imaginary
part) \cite{bib:Falkovsky-cond}. This corresponds \cite{bib:AF-prb}
to the absorption coefficient $g=4\pi\sigma/c$ which gives $g_{1}=\pi e^{2}/\hbar c\approx2.5\%$.
For the bilayer, there are four possible inter-band transitions, reflected
by its conductivity,\begin{multline}
\sigma_{2}=\frac{e^{2}}{2\hbar}\left(\frac{1}{2}\frac{\Omega+2}{\Omega+1}+\frac{1}{\Omega^{2}}\theta(\Omega-1)+\frac{1}{2}\frac{\Omega-2}{\Omega-1}\theta(\Omega-2)\right)\\
+i\frac{e^{2}}{2\pi\hbar}\Bigg(\frac{\Omega}{\Omega^{2}-1}\log\Omega+\frac{2}{\Omega}-\frac{1}{\Omega^{2}}\log\left|\frac{1+\Omega}{1-\Omega}\right|\\
-\frac{1}{2}\frac{\Omega^{2}-2}{\Omega^{2}-1}\log\left|\frac{2+\Omega}{2-\Omega}\right|-\frac{1}{2}\frac{\Omega}{\Omega^{2}-1}\log\left|4-\Omega^{2}\right|\Bigg).\label{eq:sigma2}\end{multline}
 Here $\Omega=\hbar\omega/\gamma_{1}$ is the frequency written in
units of the inter-layer coupling and $\theta(x)=[1+\mathrm{sgn}(x)]/2$.
The real part of this function has a discontinuity at $\hbar\omega=\gamma_{1}\approx0.4$eV
and a cusp at $\hbar\omega=2\gamma_{1}$. These correspond to the
activation (at zero temperature) of the interband transitions between
low-energy bands and split band, and the two split bands respectively.
The imaginary part of $\sigma_{2}$ shows a divergency at $\hbar\omega=\gamma_{1}$,
leading to an enhanced reflectance of the bilayer at this frequency.

For non-polarized light arriving at the incidence angle $\alpha$
to the sample depicted on the right-hand side of Fig. \ref{fig:exp_setup}
with graphene deposited on the top surface, the reflectance is \begin{multline}
R=\frac{1}{2}\left|\frac{\sqrt{\varepsilon_{s}}\cos\alpha_{s}D-(\cos\alpha-\frac{4\pi\sigma}{c})C}{\sqrt{\varepsilon_{s}}\cos\alpha_{s}D+(\cos\alpha+\tfrac{4\pi\sigma}{c})C}\right|^{2}\\
+\frac{1}{2}\left|\frac{\sqrt{\varepsilon_{s}}\cos\alpha C'-\cos\alpha_{s}D'(1-\frac{4\pi\sigma}{c}\cos\alpha)}{\sqrt{\varepsilon_{s}}\cos\alpha C'+\cos\alpha_{s}D'(1+\frac{4\pi\sigma}{c}\cos\alpha)}\right|^{2};\label{eq:Rfull}\end{multline}
In this result the first term represents reflection of radiation
polarized so that the electric field is perpendicular to the plane of
incidence, the second term to radiation polarised so that the electric
field is parallel to the plane of incidence, and
\begin{gather*}
A=-\sqrt{\varepsilon_{d}}\cos\alpha_{d}\cos X_{d}+i\sqrt{\varepsilon_{b}}\cos\alpha_{b}\sin X_{d},\\
B=i\sqrt{\varepsilon_{d}}\cos\alpha_{d}\cos X_{d}-\sqrt{\varepsilon_{b}}\cos\alpha_{b}\cos X_{d},\\
C=-i\sqrt{\varepsilon_{d}}\cos\alpha_{d}B\sin X_{s}+\sqrt{\varepsilon_{s}}\cos\alpha_{s}A\cos X_{s},\\
D=\sqrt{\varepsilon_{d}}\cos\alpha_{d}B\cos X_{s}-i\sqrt{\varepsilon_{s}}\cos\alpha_{s}A\sin X_{s};\end{gather*}
\begin{gather*}
A'=\sqrt{\varepsilon_{b}}\cos\alpha_{d}\cos X_{d}-i\sqrt{\varepsilon_{d}}\cos\alpha_{b}\sin X_{d},\\
B'=\sqrt{\varepsilon_{d}}\cos\alpha_{b}\cos X_{d}-i\sqrt{\varepsilon_{b}}\cos\alpha_{d}\sin X_{d},\\
C'=\sqrt{\varepsilon_{d}}\cos\alpha_{s}A'\cos X_{s}-i\sqrt{\varepsilon_{s}}\cos\alpha_{d}B'\sin X_{s},\\
D'=-i\sqrt{\varepsilon_{d}}\cos\alpha_{s}A'\sin X_{s}+\sqrt{\varepsilon_{s}}\cos\alpha_{d}B'\cos X_{s}.\end{gather*}
Here $X_{s}=\sqrt{\varepsilon_{s}}ks\cos\alpha_{s}$, $X_{d}=\sqrt{\varepsilon_{d}}kd\cos\alpha_{d}$,
$\sin\alpha_{b}=\sin\alpha/\sqrt{\varepsilon_{b}}$, $\sin\alpha_{s}=\sin\alpha/\sqrt{\varepsilon_{s}}$
and $\sin\alpha_{d}=\sin\alpha/\sqrt{\varepsilon_{d}}$. The and $\alpha$
is determined by the direction of the wave vector of the incident plane
wave, see Fig. \ref{fig:exp_setup}. 
To model a finite slab of silicon of width $d$ with a silicon oxide
layer of width $s$ on top, we substitute
$\varepsilon_{d}=\varepsilon_{\mathrm{Si}}(\omega)$,
$\varepsilon_{s}=\varepsilon_{\mathrm{SiO}_{2}}$, $\varepsilon_{b}=1$,
and the quantity $R_{0}$ is found by replacing $\sigma=0$ in these
expressions. To evaluate the visibility $V_{R}$, the integral in Eq.
(\ref{eq:R-integral}) must be taken for $R$ and $R_{0}$ using Eq.
(\ref{eq:Rfull}).

\begin{figure}[tbh]
\centering
\includegraphics[viewport=0 0 678 450, scale=0.36]{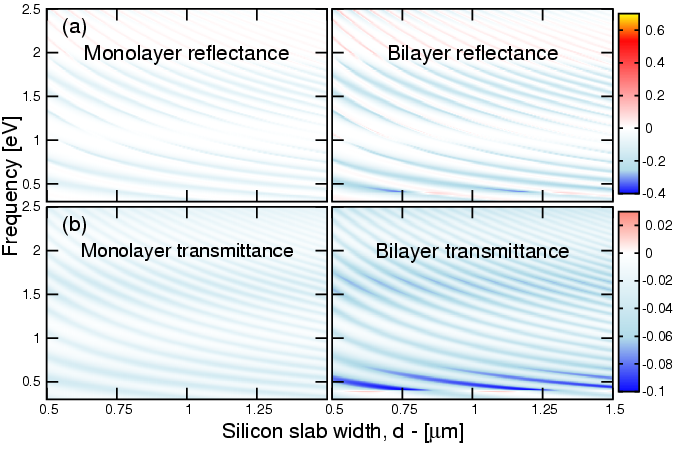}
\caption{Visibility in (a) reflectance $V_{R}$ and (b) transmittance $V_{T}$
for graphene on a silicon wafer with a 300nm oxide layer for varying
substrate thickness and frequency of radiation. Note the difference
in the scales of (a) and (b). Here we take an aperture of $\delta\alpha=10^{\circ}$
and $\bar{\alpha}=20^{\circ}$. \label{fig:Sislab-RT}}
\end{figure}

Figure \ref{fig:Sislab-RT} illustrates the visibility of mono- and
bilayer flakes on a Si substrate of widths $0.5\mu\text{m}<d<1.5\mu\text{m}$
and a 300nm SiO$_{2}$ layer (see Fig. \ref{fig:exp_setup}) for light
with $0.3\text{eV}<\hbar\omega<2.5\text{eV}$ arriving with aperture
$\delta\alpha=10^{\circ}$ around $\bar{\alpha}=20^{\circ}$. The
rapid oscillations of the visibility in this plot are caused by the
resonant condition of the Si layer. When this layer is strongly transmitting
(that is, when $\cos X_{d}\approx0$), the visibility is at its highest.
This fine structure is modulated by the corresponding resonance condition
in the oxide which is responsible for the `bands' which lie across
the plots in Fig. \ref{fig:Sislab-RT}. The condition for maximum
transmission through the oxide is $\cos X_{s}\approx0$ which leads
to \begin{equation}
\omega\approx c(n+\tfrac{1}{2})\pi/\left(s\sqrt{\varepsilon_{s}}\cos\alpha_{s}\right)\label{eq:Rmin-slab}\end{equation}
where $n$ is an integer. The wave vector of the light in the slab
is of the order of an inverse micron, so the resonant conditions are
closely spaced on the length scale of the substrate thickness. The
visibility of a bilayer flake is higher than the visibility of a monolayer
for $\hbar\omega>\gamma_{1}\approx0.4$eV because the conductivity
of the bilayer is essentially twice as large as the conductivity of
the monolayer in this energy range. Additionally, the divergency in
the imaginary part of the bilayer conductivity at $\hbar\omega=\gamma_{1}\approx0.4$eV
causes a stronger reflection and hence a larger visibility. Also we
have calculated the transmittance $T$ of the sample, and the corresponding
visiblity $V_{T}=(T-T_{0})/T_{0}$ is shown in Fig. \ref{fig:Sislab-RT}(b)
where the same resonant structure appears, but is at least ten times
weaker than the visibility in reflectance.

\begin{figure}[t]
\centering
\includegraphics[viewport=0 0 673 467, scale=0.36]{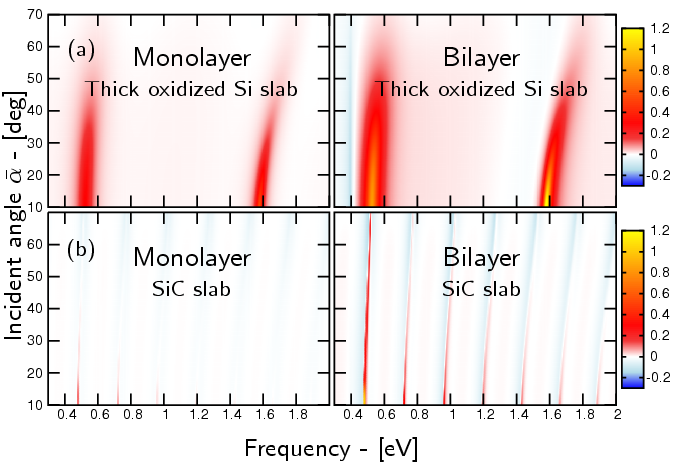}
\caption{Frequency dependence of visibility $V_{R}$ of graphene on (a) an
infinite silicon slab with a thin oxide layer of width 300nm, and
(b) a silicon carbide slab of width 1$\mu$m. In both plots we use
aperture $\delta\alpha=10^{\circ}$. \label{fig:infdi-O2}}
\end{figure}

We find that the visibility of graphene in reflectance is futher enhanced
by using a thick (semi-infinite) substrate with a sizeable oxide layer
on its surface, in agreement with a recent experimental observation
\cite{bib:blakes-ssc}. Figure \ref{fig:infdi-O2}(a) shows the visibility
of graphene deposited on a semi-infinite slab of silicon \cite{bib:permittivity-details}
with a 300nm SiO$_{2}$ layer. In this case the analytical expression
for the reflectance of a plane wave with wave vector $\vec{k}=\frac{\omega}{c}(\sin\alpha,0,-\cos\alpha)$
can be found by substituting $\varepsilon_{b}=\varepsilon_{\mathrm{Si}}(\omega)$,
$\varepsilon_{s}=\varepsilon_{\mathrm{SiO}_{2}}$ and $d=0$ into
Eq (\ref{eq:Rfull}). As before, both $R(\vec{k})$ and $R_{0}(\vec{k})$
(which is determined from this equation with $\sigma=0$), must be
substituted in Eq. (\ref{eq:R-integral}) before the visibility is
evaluated. In the plots in Fig. \ref{fig:infdi-O2}(a), the main features
are the very strong reflectance of the graphene flake at $\hbar\omega\approx0.5$eV
and $\hbar\omega\approx1.6$eV. These are due to the standing wave
resonances in the oxide layer at the condition in Eq. (\ref{eq:Rmin-slab}).
In Fig. \ref{fig:infdi-O2}(a) the peak in visibility at $\hbar\omega\approx0.5$eV,
($n=0$) corresponds to the first resonance in the oxide layer and
the peak at $\hbar\omega\approx1.6$eV ($n=1$) to the second resonance.
The factor of 2 difference between the bilayer and monolayer conductivities
at $\hbar\omega\gg\gamma_{1}$ and the divergence in the imaginary
part of $\sigma_{2}(\omega)$ at $\hbar\omega=\gamma_{1}\approx0.4$eV
are manifested in the visibility.

Besides being produced using the microcleavage technique, ultra-thin
graphitic films can also be grown by thermal annealing of SiC wafers
\cite{bib:SiC-Ohta,bib:Berger-SiC}. The reflectance for this configuration
can be found by substituting \cite{bib:permittivity-details} $d=0$,
$\varepsilon_{b}=1$ and $\varepsilon_{s}=\varepsilon_{\mathrm{SiC}}$
in Eq. (\ref{eq:Rfull}). Plots of the visibility defined by this
function are shown in Fig. \ref{fig:infdi-O2}(b). The standing wave
resonance in the substrate is again the main factor for the visibility
of graphene, though it is weaker for a SiC slab than for the SiO$_{2}$/Si
substrates.

In conclusion, we have found that graphene is much more visible in
reflection than in transmission and that the resonance condition of
the substrate is the dominating factor in determining its visibility.
For optimum visibility the wavelength of monochromatic light used
should be selected using Eq. (\ref{eq:Rmin-slab}), and for the visible
frequency range where ($\sigma_{2}\approx2\sigma_{1}$) a bilayer
is clearly distinguishable from a monolayer.


\begin{thebibliography}{14}
\bibitem{bib:permittivity-details}We take data for the real and imaginary
parts of the permittivity of Si from \cite{bib:si-permittivity}.
The imaginary part is substantial only for $\omega>1.5$eV, so we
approximate $\mathrm{Im}\epsilon_{\text{Si}}(\omega<1.5\text{eV})\approx0$.
The permittivity of SiO$_{2}$ is nearly dispersionless in this frequency
range \cite{bib:SiO2-permittivity}, so we take the SiO$_{2}$ dielectric
constant as $\varepsilon_{\mathrm{SiO}_{2}}=3.9$. SiC has some dispersion
for these frequencies \cite{bib:SiC-permittivity} but the imaginary
part is negligible so we assume $\mathrm{Im}\epsilon_{\text{SiC}}(\omega)\approx0$.
\end{thebibliography}
\end{document}